\documentclass[10pt,letterpaper]{llncs}
\usepackage{fullpage}
\usepackage{graphicx}
\usepackage{tikz}
\usetikzlibrary{bayesnet}
\usepackage{caption}

\usepackage{hyperref}
\hypersetup{
    colorlinks=true,
    linkcolor=blue,
    filecolor=magenta,
    urlcolor=cyan,
}

\usepackage{amsmath} 
\usepackage{color}
\usepackage{spverbatim}

\begin{document}

\title{SurvODE: Extrapolating Gene Expression Distribution for Early Cancer Identification}

\titlerunning{Abbreviated paper title}
%
\author{Tong Chen\inst{1} \and
Sheng Wang\inst{2}}
%
%
\institute{Institute for Interdisciplinary Information Sciences, Tsinghua University
 \\ \email{chen-t18@mails.tsinghua.edu.cn} \and
Paul G. Allen School of Computer Science \& Engineering, University of Washington \\ \email{swang@cs.washington.edu}} 
\maketitle  
\begin{abstract}
With the increasingly available large-scale cancer genomics datasets, machine learning approaches have played an important role in revealing novel insights into cancer development. Existing methods have shown encouraging performance in identifying genes that are predictive for cancer survival, but are still limited in modeling the distribution over genes. Here, we proposed a novel method that can simulate the gene expression distribution at any given time point, including those that are out of the range of the observed time points. In order to model the irregular time series where each patient is one observation, we integrated a neural ordinary differential equation (neural ODE) with cox regression into our framework. We evaluated our method on eight cancer types on TCGA and observed a substantial improvement over existing approaches. Our visualization results and further analysis indicate how our method can be used to simulate expression at the early cancer stage, offering the possibility for early cancer identification. 


\textbf{Availability}: \url{https://github.com/ct123098/SurvODE}.
\end{abstract}

\section{Introduction}
Large-scale cancer genomics projects, such as TCAG\cite{chang2013cancer}, ICGC\cite{zhang2019international} and MSK\cite{cheng2015memorial}, comprehensively characterize the genomic profiles of patients across different cancer types. Consequently, machine learning approaches have been extensively developed to analyze these datasets, revealing key insights in cancer development and drug discovery\cite{hofree2013network,tomczak2015cancer}. One important application from analyzing these datasets is to find genes whose expression can be predictive of cancer development and patient survival using the censored survival data collected in these data repositories. Methods based on Cox regression have been used to find these genes and lead to many successful downstream applications\cite{anaya2016oncolnc}.

However, these methods mainly use statistics (e.g., t-statistic, regression coefficients, etc.) to find the most significant gene, while few study model the distribution of gene expression.
Examples showing the limitations of the conventional methods are not far to seek.
If the mean of a gene in the population is less relevant to time but the variance goes up through the development of cancer, this gene is also associated with cancer, and this gene can be a potential biomarker for early detection. 
Alternatively, for some cancer, two subtypes may have different gene expression features such as high and low, but merging them together may lead to a zero regression coefficient.
Conventional methods like regression or survival analysis are unlikely to discover these patterns in the datasets. 

Our goal is to approximate the distribution of gene expression in the population whose survival time is the given number.
However, modeling the time-varying gene expression distribution is challenging due to the scarce of samples \cite{karras2020training}.
Especially, learning the distribution of gene expression in the early stage is more difficult since the number of patients with long survival times is limited.
We even have to estimate the distribution at some time that does not appear in the training set.

In this paper, we use a variational autoencoder (VAE)\cite{kingma_auto-encoding_2014} structure to model the randomness.
To better utilize the information in the datasets, we have to incorporate the data with censored survival time (we don't know the exactly surviving time of the patient and only know a lower bound), which are the majority compared to the uncensored data in the most of cancer dataset.
To explicitly capture the dynamic of gene expression distribution, we propose assumptions and formulate the problem as an irregular time series problem \cite{rubanova2019latent}, where each patient represents one observation at a given time point in a time series. 
Since the survival times of patients are unevenly spaced, they form an irregular time series that cannot be easily processed by sequence-based approaches such as recurrent neural networks (RNN)\cite{cho2014learning} and LSTM\cite{hochreiter1997long}.

We propose to model this irregular time series using neural ordinary differential equation (neural ODE)\cite{chen2018neural}, which has been used to model unevenly spaced observations in other areas, such as computer vision\cite{yildiz_ode2vae_2019}. However, it remains unclear how to model uncensored data using neural ODE as it optimizes the model using an ODE-solver that cannot be compatible with the pairwise comparison in the Cox proportional hazards (Cox-PH) module (also known as Cox regression). We address this issue by first estimating a pseudo survival time for each patient using a Cox-PH model and then constructing a time series based on these pseudo times. 

Our method will then learn the gene expression patterns for different time points and is capable of generating the gene expression at unobserved time points. Importantly, this model also enables us to extrapolate to a time point that is never observed before, offering the possibility to generate expression profiles for early cancer patients and thus find biomarkers for early cancer development. We evaluated our method on 8 cancer types that have enough survival data in TCGA and observed a substantial improvement in comparison to existing approaches. Our visualization and in-depth analysis further demonstrate how our method can be used to identify early cancer signals. We envision our method will provide novel insights into cancer analysis by generating gene expression profiles for early cancer patients that are difficult to be observed in clinics.

\section{Method}

In this section, we first introduce our problem setting. Then we introduce three key dynamics of gene expression based on our observations. We then propose SurvODE, which integrates three modules, variational autoencoder (VAE), neural ordinary differential equations (neural ODE), and a Cox proportional hazards (Cox-PH) module to jointly model the three components of the mechanism of gene expression.
Afterward, we identify three components about the dynamics of gene expression based on our assumptions.

\subsection{Problem Setting}
We want to learn the distribution of gene expression in the population given a survival time of a cancer patient.
Formally, we define the population $D = \{x_\lambda, t_\lambda\}_{\lambda \in \Lambda}$ as the observed expression data of $|\Lambda|$ patients.
Each patients $\lambda$ is associated with a gene expression vector $x_\lambda$ and a survival time $t_\lambda$.
Let $D(t) = \{x_\lambda \mid t_\lambda = t\}$ denote the gene expression of the individuals whose survival time is $t$. The survival time is often censored due to the partial observation. Therefore, the training set $T=\{x^{(i)}, u^{(i)}, \delta^{(i)}\}_{i=1}^{n}$ is collected by uniformly randomly selecting a subset of patients from the population $D$.
In $T$, the observed time $u^{(i)}$ associated with patient $i$ is censored by the experiment time $c^{(i)}$ (i.e., $u^{(i)} = \min(t^{(i)}, c^{(i)})$ if let $t^{(i)}$ denote the ground-truth survival time), and the $\delta^{(i)}$ means whether the survival time is observed (i.e., $\delta^{(i)} = \mathbf{1}\{t^{(i)} \leq c^{(i)}\}$).
Our goal is to learn from the training set $T$ and simulate the distribution of gene expression in the population $D(t)$ given arbitrary survival time $t$. 

To establish our analysis, we make the following assumptions. 
The first two assumptions provide the intuition behind the survival analysis model and neural ODE to improve the generalizability, and the third assumptions are for the Cox-PH model.
\begin{itemize}
	\item
	We assume if we know the gene expression vector $x$ of a patient, its survival time $t$ should follow a certain distribution. In population distribution $D$, conditioning on a gene expression $x$, the survival time $t$ is a random variable.
	
	\item 
	We assume the gene expression vector $x$ of patient $t$ evolves according to a hidden dynamic regarding the time. 
	The evolving dynamics of the gene expression follow the first-order ordinary differential equation in a low dimensional space.
	
	\item 
	We further make a few common assumptions in survival analysis.
	To infer ground-truth survival time from the censored survival time, we assume the experiment time $c^{(i)}$ is independent of the ground-truth survival time $t^{(i)}$.
	To utilize the Cox regression, we adopt the proportional hazards assumption.
	Given the gene expression vector $x$, the hazard ratio of survival time $h(t \mid x)$ is the multiplication of $h_0(t)$, which only depends on $t$, and $\exp(\beta^T \cdot x)$, which only depends on $x$.
	
\end{itemize}

We further assume that at any time $t$, the gene expression vector
$x$, has the hazard ratio of $\lambda(x)$. This means we may observe the event in the interval $[t, t+\varepsilon)$ with probability $\lambda(x) \cdot \varepsilon$ for a infinitesimal $\varepsilon$.

\subsection{Framework}
\begin{figure}[h]\centering
\includegraphics[width=0.95\textwidth]{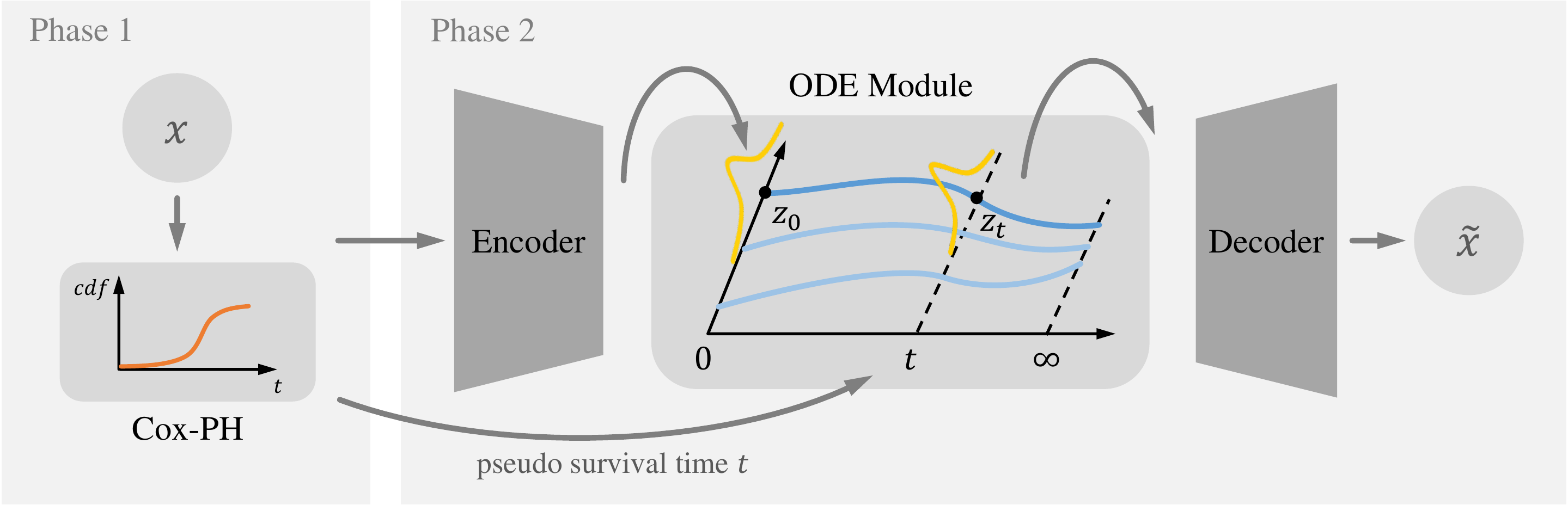}
\caption{\textbf{Overall design of SurvODE.} SurvODE consists of three components, the variational autoencoder (VAE) module (yellow), the neural ordinary differential equation (neural ODE) module (blue), and the Cox proportional hazard (Cox-PH) module (orange). During the training, we first train the Cox-PH module in phase one. In phase two, the pretrained Cox-PH module generates the distribution of survival time given a gene expression $x$. The encoder maps a gene expression and the sampled pseudo survival time to a latent representation $z_0$. The neural ODE model transits the $z_0$ to another latent representation $z_t$. And the decoder maps the $z_t$ to a reconstructed gene expression $\tilde{x}$. We train the neural networks by minimizing a reconstruction loss and some regularization terms. In evaluating, we sample $z_0$ from $N(0, 1)$ and forward it through the neural ODE and decoder and construct the gene expression.}
\label{fig:model}
\end{figure}

Different from previous works, we focus on estimating and simulating the distribution of gene expression rather than the mean.
However, estimating the distribution remains challenging due to the limited patient data. To tackle this challenge, we investigate the mechanism of gene expression in the population and decompose the mechanism into three components.
\begin{itemize}
	\item 
	The gene expression when we observe the event may vary among patients.
	\item 
    The gene expression evolves according to some unknown dynamics.
	\item 
    The survival time of a patient has randomness given its gene expression.
\end{itemize}
SurvODE consists of three modules, each of which characterizes one component and can improve the generalizability of our model.
The three components are the variational autoencoder (VAE) module (yellow), the neural ordinary differential equation (neural ODE) module (blue), and the Cox proportional hazard (Cox-PH) module (orange)
In the next few subsections, we first introduce the CVAE module, which learns a low-dimensional representation of gene expression.
The CVAE helps us to interpolate valid gene expressions which are not seen during the training. 
Secondly, we introduce the ODE module, which models the dynamics of gene expression evolving.
The ODE Neural Network learns how a patient's gene expression evolves exploits the similarity between gene expression samples in the low-dimensional space. 
Finally, we introduce a Cox-PH module, which learns the distribution of survival time given the gene expression at some time.
All three modules improve the generalizability and the prediction ability of our methods, especially in the extrapolation setting. 

\subsection{Variational Autoencoder Module}
We first introduce the variational autoencoder (VAE) module.
The VAE contains two neural networks, the encoder, and the decoder. 
The variation of our output comes from the VAE module. 

The \textbf{encoder} is a Multi-layer Perceptron (MLP), denoted as $\theta_E$.
The encoder maps the gene vector $x$ and a survival time $t$ to a normal distribution.
The normal distribution is defined by the mean and the variance. 
Therefore, the output of the encoder are the mean $\mu$ and the log-variance $s$. 
We sample the latent representation $z_0$ from the normal distribution $N(\mu, \exp(s))$, which corresponds to the gene expression when we observe an event. 
\begin{equation*}
    (\mu, s) = f_{\theta_E}(x, t), \qquad
    z_0 \sim N(\mu, \exp(s))
\end{equation*}

The \textbf{decoder} is a a MLP symmetric to the encoder, denoted as $\theta_E$.
It maps a position in the latent space $z_t$ to the reconstruct gene expression $\tilde{x}$.
$z_t$ corresponds to the gene expression of a patient whose survival time is exactly $t$.
\begin{equation*}
    \tilde{x} = f_{\theta_D}(z_t)
\end{equation*}



\subsection{Neural ODE Module}
The neural ODE module characterizes the dynamic of the progress of the disease in the latent space. 
It predicts the latent representation of a gene expression $z_t$ given the latent representation of the gene expression when the event is observed $z_0$ and the survival time $t$.

Although conventional neural network structures like MLP have the approximation ability, the generalizability is weak. 
We incorporate an advanced neural network technique called neural ODE to improve the ability to extrapolate unseen survival time $t$. 

The neural ODE defines a function by specifying its derivative using a neural network
\begin{equation*}
\frac{\partial}{\partial t} F(z, t) = f_{\theta_O}(z)
,\end{equation*}
where $f_{\theta_O}$ represent a MLP.
And thus, the the latent representation of a gene expression at time $t$ is given by
\begin{equation*}
    z_t = F(z_0, t)
.\end{equation*}
The forward and backward details about neural ODE is introduced in \cite{chen2018neural}.





\subsection{Survival Module}
After using the VAE module and neural ODE module, we can already generate the gene expression vector at any time point. 
However, there are still some unsolved difficulties. 
First of all, the uncensored data in the training set are too sparse. 
Second, during the training, only a small number of $z_t$ is associated with a large value of $t$ and such observations are very sparse and might make neural network overfit.
Therefore, the decoder does train with latent space sampled from large $t$. 
This leads to poor generalizability. 
Potentially, training with the censored data can improve the generalizability since there is more sample with large $t$.
However, how we utilize the censored data efficiently is unsure. 
We propose a generative-based component to incorporate censored data to efficiently use the data to improve generalizability.
A cox-PH model is used to ``generate'' samples for training to improve the training for large $t$.

The Cox-PH model is introduced in [xxx].
Let $\beta$ denoted the linear coefficient and $h_0(t)$ denote the baseline hazard ratio. 
Cox-PH maximize the likelihood under the model $h(t \mid x) = h_0(t) \cdot \exp(\beta^T \cdot x)$.
Moreover, the cox-PH model estimates the cumulative distribution function (c.d.f.) of the survival time given a gene expression.
Let $S_0(t)$ denote the baseline survival function, and the $S_0^{-1}$ denote its inverse function. 

For a trained Cox-PH, the distribution of survival time $t$ is determined given the gene expression $x$.
And we can sample $t$ as follows.
\begin{equation*}
    t = S_0^{-1}\left(\epsilon^{1 / \exp(\beta^T \cdot x)}\right) \text{ where } \epsilon \sim U(0, 1)
\end{equation*}
Therefore, the Cox-PH model deals with the censored data and alleviates the sparsity of large $t$ by modeling the distribution of the survival time.

\subsection{Optimization}

The model is trained in two phases.
In the first phase, we train the Cox-PH module with all data by maximizing partial likelihood function \cite{cox_regression_1972}, which is a conventional way of training the Cox-PH model.
The Cox-PH model learns the coefficient of each gene and the baseline function.
In the second phase, we train the VAE module and the neural ODE module by minimizing a lower bound of the log-likelihood function.
In each epoch, we sample a pseudo survival time from the pretrained Cox-PH module for each patient.
The gene expression $x$ and the sampled pseudo survival time are passed to the VAE module and the neural ODE module to reconstruct the gene expression.
A training loss is defined in the following to let the neural network generate distribution similar to the training data.

We use $q(\cdot)$ to represent the probability of the sample generated in our model when there is no ambiguity.
Let $x$ denote the gene expression vector, $u$ denote the censored survival time, and $\delta$ denote whether the survival time is observed. 
The log-likelihood of the gene expression and the survival time generated by our method is
\begin{equation}
    \log q(x, u, \delta) = \log q(u, \delta \mid x) + \log q(x)
.\end{equation}
Note that the first term is the objective of the cox-PH module.
In the first step, we maximize $\sum_{i=1}^{n} q(u^{(i)}, \delta^{(i)} \mid x^{(i)})$ by maximizing partial likelihood function \cite{cox_regression_1972}, which is a conventional way solving Cox-PH model.
We do not go into the details since the details are irrelevant to the mainstream of this paper.

We then derive the second term. 
The random variable $t$ represent the distribution given $x$ learned by the Cox-PH module, and the random variable $z_0$ and $z_t$ are the value forwarding by the encoder and the neural ode. 
For convenience, let $Z = (z_0, z_t)$.
\begin{center}
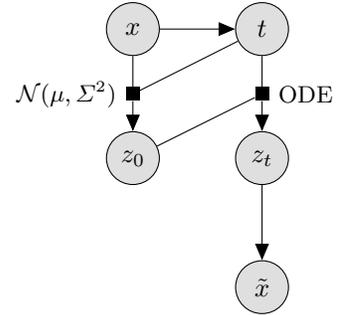

\begin{minipage}{0.69\textwidth}
\begin{equation*}
\begin{aligned}
    &{} \log q(x) 
    = E_{t} \left(\log q(x \mid t)\right) \\
    \geq& E_{t} \left( E_{Z} \left(\log \frac{q(Z \mid t, x)}{p(Z \mid t)} + \log q(x \mid t, Z)\right) \right) \\
    =& E_{t} \left( E_{Z} \left(\log \frac{q(z_0 \mid t, x) \cdot q(z_t \mid t, x, z_0)}{p(z_0 \mid t) \cdot p(z_t \mid t, z_0)} + \log q(x \mid t, z_0, z_t)\right) \right) \\
    \approx& E_{t, Z} \left(
        \underbrace{\log \frac{q(z_0 \mid t, x)}{p(z_0)}}_\text{$L_{kld_0}$} + 
        \underbrace{\log \frac{q(z_t \mid x, t)}{p(z_t)}}_\text{$L_{kld_t}$} + 
        \underbrace{\|x - \tilde{x}\|_2^2}_\text{$L_{rec}$}
    \right)
\end{aligned}
\end{equation*}
\end{minipage}
\begin{minipage}{0.29\textwidth}\centering
    \begin{tikzpicture}
    \node[obs]      (x) {$x$};
    \node[obs, right=of x]      (t) {$t$};
    \node[obs, below=of x]      (z0) {$z_0$};
    \node[obs, below=of t]      (zt) {$z_t$};
    \node[obs, below=of zt]      (xg) {$\tilde{x}$};
    
    \factor[above=of z0] {z0-f} {left:$\mathcal{N}(\mu, \Sigma^2)$} {x,t} {z0} ;
    \factor[above=of zt] {zt-f} {right: ODE} {z0,t} {zt} ; %

    \edge {x} {t} ; %
    \edge {zt} {xg} ; %
    \end{tikzpicture}
    \captionof{figure}{\textbf{Probability graphical model of SurvODE}} 
    \label{fig:pgm}
\end{minipage}
\end{center}


In a single forwarding step, we input $x$ and let
\begin{equation*}
    L_{rec}(x) = \|x - \tilde{x}\|_2^2, \quad
    L_{kld_0}(x) = \log \frac{q(z_0 \mid x)}{p(z_0)}, \quad
    L_{kld_t}(x) = \log \frac{q(z_t \mid t, z_0)}{p(z_t)}
,\end{equation*}
where $L_{rec}$ is the reconstruction loss, $L_{kld_0}$ is the KL-divergence of initial latent position, $L_{kld_t}$ is the KL-divergence of the third latent position, $p(\cdot)$ is the probability of a standard multivariate normal distribution.  
As a commonly used trick in VAE, we use $\beta_{kld_0}$ and $\beta_{kld_t}$ as two weights to balance the terms in the loss.
Therefore, the training loss of an epoch is
\begin{equation*}
    L = \sum_{i=1}^{n} L_{rec}(x^{(i)}) + \beta_{kld_0} \cdot L_{kld_0}(x^{(i)}) + \beta_{kld_t} \cdot L_{kld_t}(x^{(i)})
.\end{equation*}
We use stochastic gradient descent (SGD) to optimize the neural networks.

\section{Experiments Setup}

\subsection{Dataset}

We compared our method with baselines on datasets from The Cancer Genome Atlas (TCGA).
We only used datasets that have enough patients with uncensored survival time since we need to compare the distribution of ground-truth samples and the distribution of the generative models.
Eight cancer datasets where the uncensored samples are more than 100 are selected: bladder carcinoma (BLCA), breast carcinoma (BRCA), glioma (GBMLGG), head and neck squamous cell carcinoma (HNSC), kidney renal cell carcinoma (KIRC), brain lower-grade glioma (LGG), lung adenocarcinoma (LUAD), ovarian carcinoma (OV).

For each dataset, preprocessing was done as follows.
We removed a gene if the gene expression is not zero in more than 15\% patients. 
We scaled each gene to zero mean and one variance. 
To avoid outliers, we further removed genes that contain an outlier larger than a threshold 6. 

Since most genes are less relevant to time, we selected 100 most time-relevant genes for each dataset in our experiment.
To do this, we calculated a ``time-vary score'' for each gene. 
For each dataset, we equally split the patients with uncensored survival time into two groups.
The first group contained patients with smaller survival times and the second group contained the remaining. 
For each gene, we calculated the Wasserstein distance of the gene expression between the two groups. 
We selected the 100 genes with the top ``time-vary score'' in our experiments. 

\subsection{Metrics}
We applied the Wasserstein distance\cite{villani2009wasserstein} to measure the similarity of two distributions.
The Wasserstein distance is a widely used metric in generative problems. \cite{alaa_how_2021} 
Some other statistical divergence metrics were not used due to different reasons. 
KL divergence is not symmetric, and more importantly, it can be infinity when the supports of the two distributions are different.
The likelihood is also not suitable since it is model-dependent.

For evaluation, we split the test set into several bins according to the ground-truth survival time. 
For each bin, we used the mean survival time as a representative. 
And we generated a set of gene expression vectors from the generative model given the representative time. 
We compared the Wasserstein distance between the generated set and the ground-truth set of gene expression.


\subsection{Training Details}

In phase one, we trained the Cox-PH module.
For the Cox-PH module, we set the $l_1$ regularization to be 0.1 and the number of maximum iteration to be 1000.

In phase two, we fixed the Cox-PH module and trained the neural networks.
Let the number of genes be 100 and the dimension of the latent space be 4.
For the encoder and the decoder, we used a multiple-layer perceptron (MLP) with two layers and 32 neurons in each layer, and let the activation function be ReLU.
For the neural ODE module, we used an MLP with two layers and 32 neurons in each layer and let the activation function be Tanh. 
To balance the reconstruct loss and the KL divergence, we set the $\beta_{kld_0} = \beta_{kld_t} = 0.5 \cdot \texttt{\#gene}/\texttt{\#dim of latent} = 0.02$. 
We trained the model by the Adam optimizer with a learning rate of $10^{-3}$.



\section{Results}

In this section, experiments results are shown to illustrate our advantage over baseline methods.
In the extrapolation task, SurvODE outperforms the baseline methods VAE and CVAE, indicating that our design improves the generalizability. 
We further showed by an ablation study each of the components improves the performance of SurvODE.
We analyzed how SurvODE works by visualization, showing that the latent space is informative and the ODE module learns the trend of cancer development. 
Finally, we accomplished discovering cancer biology knowledge using SurvODE. 
In GBMLGG, we found that most top-ranked genes are associated with literature evidence.

\subsection{Improved Extrapolation Ability in TCGA Dataset}

\begin{figure}[h]\centering
\includegraphics[width=0.8\textwidth]{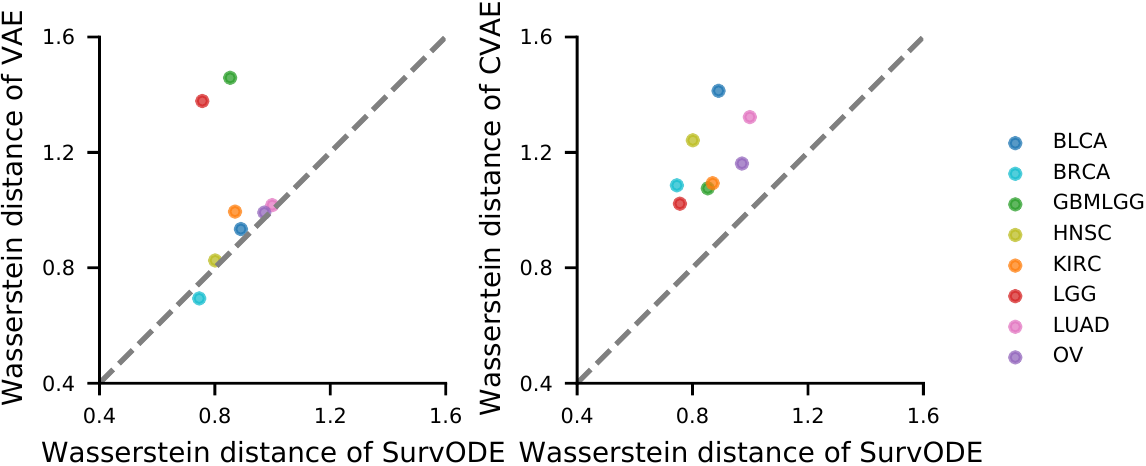}
\caption{\textbf{Scatter plots comparing the performance between SurvODE and the comparison approaches in extrapolation task in terms of the Wasserstein distance.} Compared with VAE (left), SurvODE has better performance, indicating SurvODE can extrapolate gene distribution at unseen time points. Also, SurvODE outperforms the CVAE (right), indicating SurvODE has better generalizability, especially in the extrapolation task.}
\label{fig:comp}
\end{figure}

We first sought to evaluate the performance of SurvODE in the extrapolation task (Fig. \ref{fig:comp}). 
For each dataset, we split 20\% uncensored patients with the largest ground-truth survival time as the test set.
The test set is further split into two bins by the survival time.
The remaining patients are in the training set. 

To show our model generates meaningful early-stage gene expression, we compare our model with VAE, which is a model generating the same distribution ignoring time.
VAE is obtained by removing the neural ODE and the cox-PH module from SurvODE. 
When compared to the VAE, our method obtains improvements on all cancers except one.
Notably, the improvement is rather significant in GBMLGG and LGG. 
However, in BRCA, the VAE is better than our model. 
It is because the gene expression distribution at the unseen time points is different from the training samples, which leads to a bad performance of all models containing the time as input. 
We also applied the paired rank-sum test showing it is likely that our model tends to have better performance than VAE in extrapolation tasks (p-value = 0.115). 
This result indicating using a generative model to extrapolate gene expression at the early stage of cancer is feasible.

To handle generating samples given labels, A widely used model is CVAE.
The difference between CVAE and VAE is that the encoder and the decoder have the time as their input and generate different distributions given different survival times.
We compared our method to the CVAE and observed a superior performance of SurvODE.
In all eight cancer datasets, SurvODE outperforms CVAE with a significant amount, confirming the effectiveness of the neural ODE module and Cox-PH module. 
Also, applying the paired rank-sum test shows that the improvement is significant (p-value = 0.001).
Thus, SurvODE is able to extrapolates gene expression at the early stage of cancer

\subsection{Ablation Study}

\begin{figure}[h]\centering
\includegraphics[width=0.8\textwidth]{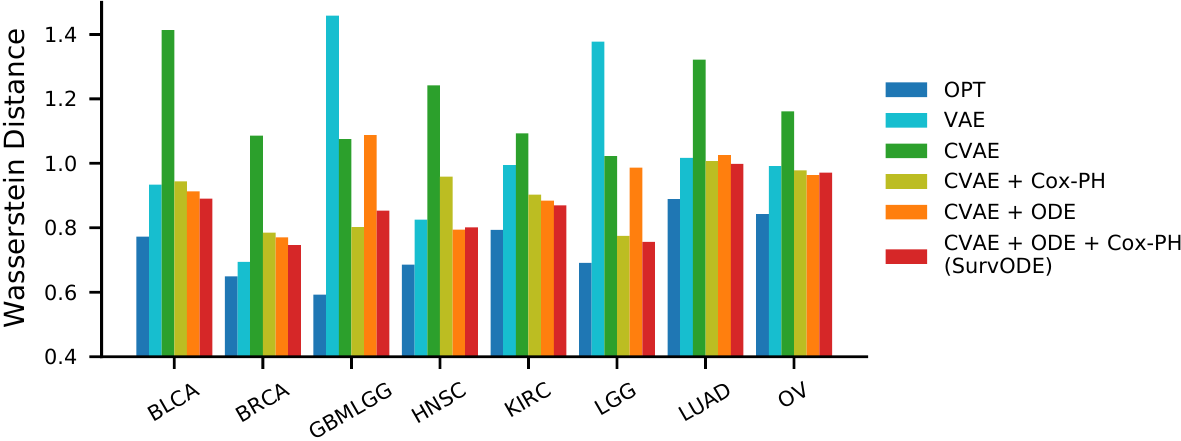}
\caption{\textbf{Performance Comparison of Extrapolation Task in Cancer Datasets.} SurvODE has a better performance compared with baselines and methods using only one module. }
\label{fig:extra}
\end{figure}

We evaluate some comparison method in the eight cancer data in extrapolation task.
OPT is a CVAE training with all data, which is an empirical lower bound of using VAE-based method.
CVAE + Cox-PH is a simplified SurvODE where we use a MLP instead of neural ode module.
CVAE + ODE is another simplified SurvODE where we do not use Cox-PH module and censored data.

We observed that the performance of CVAE is even worse than VAE in most datasets (BLCA, BRCA, HNSC, KIRC, LUAD, and OV), indicating CVAE easily overfits the training set and cannot correctly simulate the gene expression distribution at the time points which does not appear during training. (Fig. \ref{fig:extra})
While SurvODE alleviates the overfitting problem and successfully extrapolates gene expression at the early stage of cancer.

We further sought to analyze the necessity of each component in our method to improve the generalizability. 
When compared to the CVAE without Cox-PH, CVAE with Cox-PH obtains improvements on all cancers.
When compared to the CVAE without ODE module, CVAE with ODE module obtains improvements on all cancers except one (GBMLGG).
Comparing CVAE with Cox-PH and CVAE with ODE module to SurvODE, SurvODE has the best performance among them in five datasets and is slightly worse than the top one in the other three datasets.
This implies SurvODE has better generalizability than only using one module.

\subsection{Interpolation Task}

\begin{figure}[h]\centering
\includegraphics[width=0.8\textwidth]{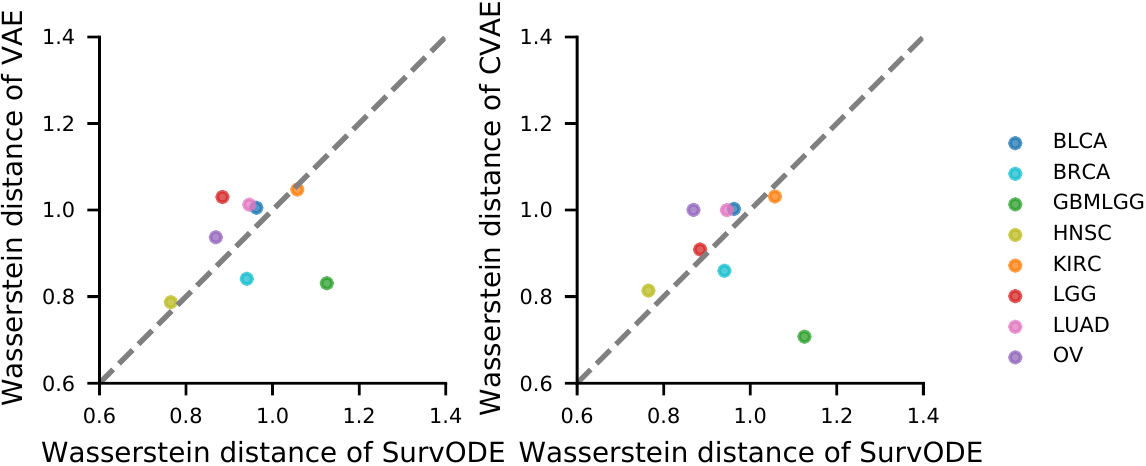}
\caption{\textbf{Performance Comparison of Interpolation Task in Cancer Datasets}. SurvODE achieves comparable performance in the interpolating task, indicating that SurvODE has the same fitting ability as VAE and CVAE. This implies the excellent generalizability does not impair the fitting ability.}
\label{fig:inter}
\end{figure}
After verifying the excellent performance of our method in the extrapolation task, we next examine whether the neural ODE module and Cox-PH module impair the performance in the ``validation set''. (Fig. \ref{fig:inter})
The validation set here means we leave out some samples from the training set, and we test the performance by evaluating the similarity between the generated of our method and the leave-out samples.

To model the variation of the gene expression, we incorporate the ODE module and Cox-PH model to improve the generalizability of the neural networks.
ODE module can be regarded as a strong regularization compared with common neural networks.
Moreover, the Cox-PH model may introduce additional errors since it is based on proportional hazard assumptions and linear approximation.
The improvement of extrapolation ability might compensate for the fitting accuracy within the training distribution.
However, we want to show that our model can also achieve the same performance with baseline within the training distribution despite better extrapolation ability (Fig. \ref{fig:inter}).

When compared to the VAE, our method even obtains improvements in the five datasets (BRCA, HNSC, LGG, LUAD, and OV). 
We also compared our model to the CVAE and observed a similar result, improvements in the five datasets (BRCA, HNSC, LGG, LUAD, and OV). 
Although in BRCA and KIRC, our model is slightly worse than baselines in Wasserstein distance, our model and baselines are comparable.
However, in GBMLGG, our model is much worse than VAE and CVAE. 
The reason might be the error introduced by the Cox-PH model.

\subsection{Visualization}

\begin{figure}[h]\centering
\begin{minipage}{0.45\textwidth}\centering
\includegraphics[width=0.99\textwidth]{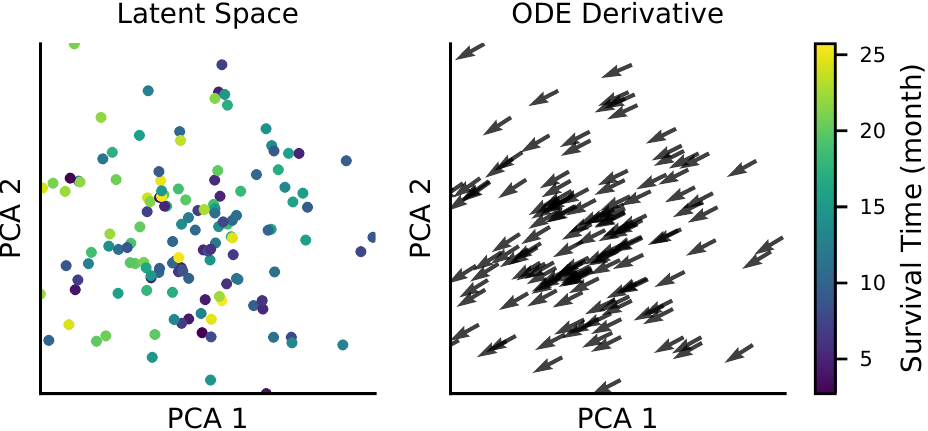}
(a) BLCA
\end{minipage}
\begin{minipage}{0.45\textwidth}\centering
\includegraphics[width=0.99\textwidth]{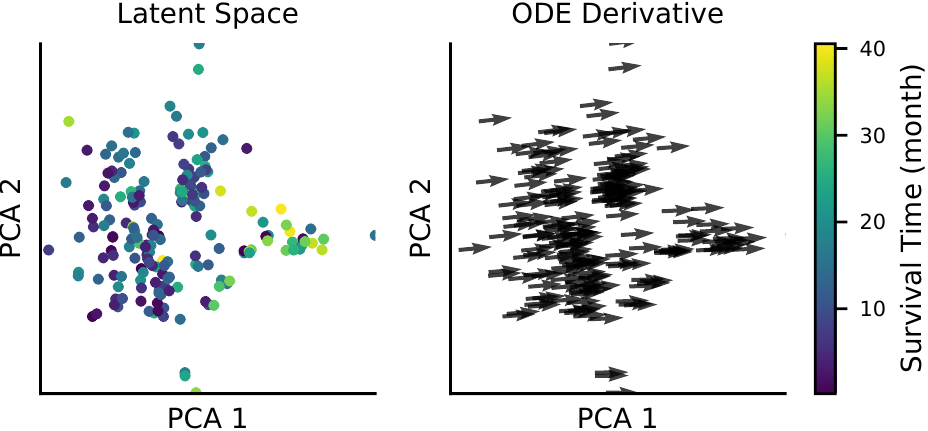}
(b) GBMLGG
\end{minipage}
\par
\begin{minipage}{0.45\textwidth}\centering
\includegraphics[width=0.99\textwidth]{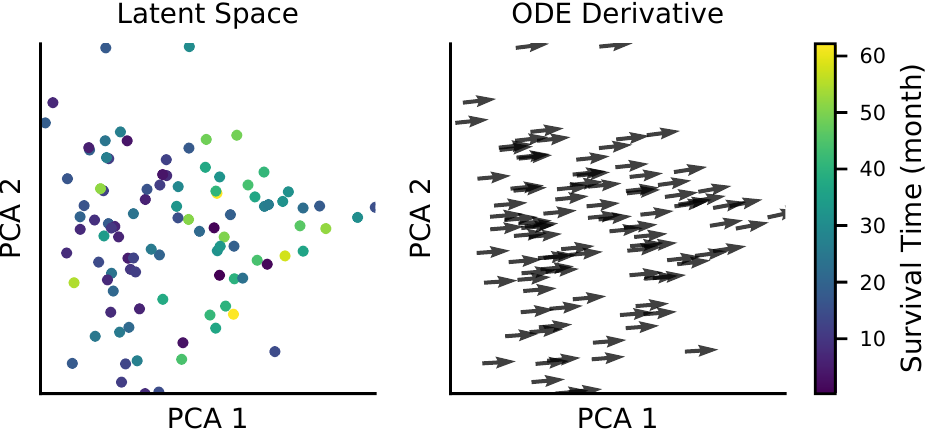}
(c) LGG
\end{minipage}
\begin{minipage}{0.45\textwidth}\centering
\includegraphics[width=0.99\textwidth]{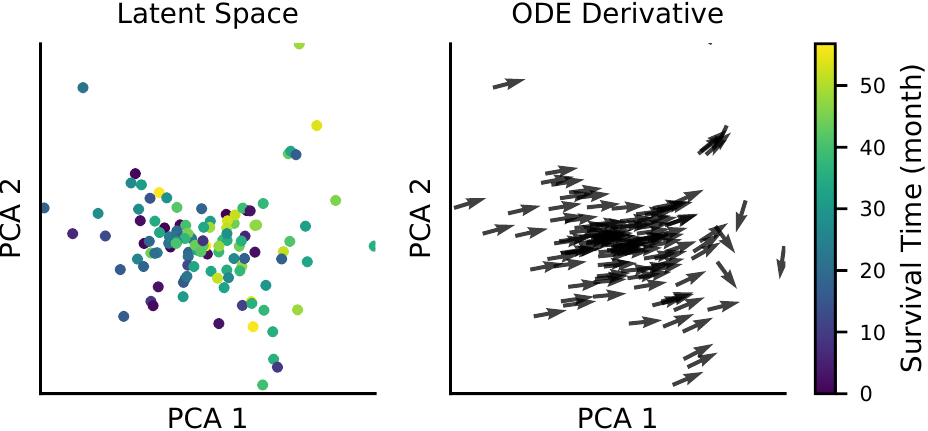}
(d) OV
\end{minipage}
\caption{\textbf{Visualization of the Latent Space of SurvODE.} (a-d) The color of points have trends in the latent space plots (left), indicating SurvODE learns a meaningful latent space representation. In the ODE derivative plots (right), the directions of arrows point from the area of long survival time (yellow) to the area of short survival time (purple), indicating the neural ODE module also learns the correct patterns.}
\label{fig:vis}
\end{figure}

We sought to investigate the excellent extrapolation performance of our model. 
We show that the hidden space captures meaningful information to recovery the gene expression.
Especially, we show the latent representation contains much information about survival time, which is crucial for our model to generate time-vary gene expression distribution.

The process of visualizing the latent space is similar to the process of training.
We select all patients whose survival time is observed.
We randomly sample a pseudo survival time in the Cox-PH module and forward the gene expression and pseudo survival time in the model.
We query the value of $z_t$ in the model, and plot $z_t$ on a two-dimensional canvas by Principal Component Analysis (PCA) (left in Fig. \ref{fig:vis}).
additional, we plot the derivation of the neural ODE module at $z_t$ (right in Fig. \ref{fig:vis}).

We demonstrate the plots of four cancers.
We can observe that for each of dataset, there is a strong correlation between the ground-truth survival time and some direction in the latent space. 
The points with high survival time are located on one side and points with low survival time are located on another side.
This implies the latent space contains much information about survival time. 

Moreover, we observe the arrows in the derivative plot pointing from a location with low survival time to one of high survival time.
The direction is consistent with the dynamics of gene expression evolving.
Thus, this implies the ODE module learns meaningful information as well.


\subsection{Genes for Early Detection}

Motivated by the improved performance of SurvODE in extrapolating gene expression distribution, we then examined which genes are likely to be a predictor of cancer development.

Specifically, we train SurvODE with all data.
We first determine the range of survival time that we are interested in.
We split the range of survival time into many bins. 
From SurvODE, we generated many gene expression samples given the meantime of each bin. 
We let the bin with the largest survival time as a target, which can be regarded as the gene expression distribution in normal people.
We compare the distribution of each gene in each bin with the distribution of the same gene in the target bin. 
For each bin, we rank the gene with the Wasserstein distance between the current bin and the target bin.
The genes with top rank are likely to be the indicator of cancer progression.
    
In GBMLGG, we split 20 bins along the survival time, count the occurrence of each gene becoming top 5 in each bin, and keep the gene if the occurrence is no smaller than 2. 
We obtained 16 genes in total:
TAGLN2, IGFBP2, EMP3, TIMP1, GPX8, RAB34, ADAM12, NSUN7,
SH2D4A, CHI3L1, PLAT, VAV3, FBXO17, SIRT1, FAM20C, and RARRES2.
We found that some of these genes can be supported by literature evidence \cite{lindstrom2019expanding,yue2018high,aaberg2018co,kodama2004adam12,sato2020five,steponaitis2016high,salhia2008guanine,li2019effects,du2018clinical,du2020secretory}. For example, TAGLN2 has been reported to be a biomarker of gliomas and promote tumorigenesis \cite{han2017tagln2}. Although we only show an example of using the distribution of individual genes, our model supports the analysis of gene pairs or even gene tuples. 
We can compute the distributional difference of gene pairs (or tuples) between bins, which might lead to the identification of more complex interactions between genes. 

\section{Conclusion}
In this paper, we proposed SurvODE, a novel method simulating the gene expression distribution at any given time point, including those that are out of the range of the observed time points.
We integrated a neural ODE module and a Cox-PH module into our framework to model the irregular time series where each patient is one observation in the time series.
We evaluated our method on eight cancer types on TCGA and observed a substantial improvement over existing approaches.
Our visualization results and further analysis indicate how our method can be used to simulate expression at the early cancer stage, offering a possibility for early cancer identification. 

In the future, we hope to test more cancer datasets to further evaluate the performance of SurvODE and make more discoveries of early cancer identification.
A limitation of our study is that we only test eight cancers, while some of them does not have a significant difference in distribution at different time point.
The reason why we did not test other cancer in the TGCA is that they do not contain enough uncensored data and ground-truth survival time to validate our method.
More experiments with a larger number of genes can be done if additional high-quality datasets are available.
On the other hand, SurvODE makes some assumptions of the underlining mechanics of gene expression.
For example, the proportional hazard assumption assumes a linear relationship between the logarithm of hazard ratio and gene expression.
This assumption harms the performance in some cancer when the assumption does not hold. 
A design with fewer assumptions may further improve the performance. 

\bibliographystyle{splncs04}
\bibliography{Genegen}

\begin{thebibliography}{10}
\providecommand{\url}[1]{\texttt{#1}}
\providecommand{\urlprefix}{URL }
\providecommand{\doi}[1]{https://doi.org/#1}

\bibitem{aaberg2018co}
Aaberg-Jessen, C., S{\o}rensen, M.D., Matos, A.L., Moreira, J.M., Br{\"u}nner,
  N., Knudsen, A., Kristensen, B.W.: Co-expression of timp-1 and its cell
  surface binding partner cd63 in glioblastomas. BMC cancer  \textbf{18}(1),
  1--16 (2018)

\bibitem{alaa_how_2021}
Alaa, A.M., van Breugel, B., Saveliev, E., van~der Schaar, M.: How {Faithful}
  is your {Synthetic} {Data}? {Sample}-level {Metrics} for {Evaluating} and
  {Auditing} {Generative} {Models}. arXiv:2102.08921 [cs, stat]  (Feb 2021),
  \url{http://arxiv.org/abs/2102.08921}, arXiv: 2102.08921

\bibitem{anaya2016oncolnc}
Anaya, J.: Oncolnc: linking tcga survival data to mrnas, mirnas, and lncrnas.
  PeerJ Computer Science  \textbf{2}, ~e67 (2016)

\bibitem{chang2013cancer}
Chang, K., Creighton, C., Davis, C., Donehower, L., et~al.: The cancer genome
  atlas pan-cancer analysis project. Nat Genet  \textbf{45}(10),  1113--1120
  (2013)

\bibitem{chen2018neural}
Chen, R.T., Rubanova, Y., Bettencourt, J., Duvenaud, D.: Neural ordinary
  differential equations. In: Proceedings of the 32nd International Conference
  on Neural Information Processing Systems. pp. 6572--6583 (2018)

\bibitem{cheng2015memorial}
Cheng, D.T., Mitchell, T.N., Zehir, A., Shah, R.H., Benayed, R., Syed, A.,
  Chandramohan, R., Liu, Z.Y., Won, H.H., Scott, S.N., et~al.: Memorial sloan
  kettering-integrated mutation profiling of actionable cancer targets
  (msk-impact): a hybridization capture-based next-generation sequencing
  clinical assay for solid tumor molecular oncology. The Journal of molecular
  diagnostics  \textbf{17}(3),  251--264 (2015)

\bibitem{cho2014learning}
Cho, K., Van~Merri{\"e}nboer, B., Gulcehre, C., Bahdanau, D., Bougares, F.,
  Schwenk, H., Bengio, Y.: Learning phrase representations using rnn
  encoder-decoder for statistical machine translation. arXiv preprint
  arXiv:1406.1078  (2014)

\bibitem{cox_regression_1972}
Cox, D.R.: Regression {Models} and {Life}-{Tables}. Journal of the Royal
  Statistical Society: Series B (Methodological)  \textbf{34}(2),  187--202
  (1972). \doi{10.1111/j.2517-6161.1972.tb00899.x},
  \url{https://onlinelibrary.wiley.com/doi/abs/10.1111/j.2517-6161.1972.tb00899.x},
  \_eprint:
  https://onlinelibrary.wiley.com/doi/pdf/10.1111/j.2517-6161.1972.tb00899.x

\bibitem{du2018clinical}
Du, D., Yuan, J., Ma, W., Ning, J., Weinstein, J.N., Yuan, X., Fuller, G.N.,
  Liu, Y.: Clinical significance of fbxo17 gene expression in high-grade
  glioma. BMC cancer  \textbf{18}(1),  1--10 (2018)

\bibitem{du2020secretory}
Du, S., Guan, S., Zhu, C., Guo, Q., Cao, J., Guan, G., Cheng, W., Cheng, P.,
  Wu, A.: Secretory pathway kinase fam20c, a marker for glioma invasion and
  malignancy, predicts poor prognosis of glioma. OncoTargets and therapy
  \textbf{13},  11755 (2020)

\bibitem{han2017tagln2}
Han, M.Z., Xu, R., Xu, Y.Y., Zhang, X., Ni, S.L., Huang, B., Chen, A.J., Wei,
  Y.Z., Wang, S., Li, W.J., et~al.: Tagln2 is a candidate prognostic biomarker
  promoting tumorigenesis in human gliomas. Journal of Experimental \& Clinical
  Cancer Research  \textbf{36}(1),  1--14 (2017)

\bibitem{hochreiter1997long}
Hochreiter, S., Schmidhuber, J.: Long short-term memory. Neural computation
  \textbf{9}(8),  1735--1780 (1997)

\bibitem{hofree2013network}
Hofree, M., Shen, J.P., Carter, H., Gross, A., Ideker, T.: Network-based
  stratification of tumor mutations. Nature methods  \textbf{10}(11),
  1108--1115 (2013)

\bibitem{karras2020training}
Karras, T., Aittala, M., Hellsten, J., Laine, S., Lehtinen, J., Aila, T.:
  Training generative adversarial networks with limited data. arXiv preprint
  arXiv:2006.06676  (2020)

\bibitem{kingma_auto-encoding_2014}
Kingma, D.P., Welling, M.: Auto-{Encoding} {Variational} {Bayes}.
  arXiv:1312.6114 [cs, stat]  (May 2014), \url{http://arxiv.org/abs/1312.6114},
  arXiv: 1312.6114

\bibitem{kodama2004adam12}
Kodama, T., Ikeda, E., Okada, A., Ohtsuka, T., Shimoda, M., Shiomi, T.,
  Yoshida, K., Nakada, M., Ohuchi, E., Okada, Y.: Adam12 is selectively
  overexpressed in human glioblastomas and is associated with glioblastoma cell
  proliferation and shedding of heparin-binding epidermal growth factor. The
  American journal of pathology  \textbf{165}(5),  1743--1753 (2004)

\bibitem{li2019effects}
Li, Y., Chen, X., Cui, Y., Wei, Q., Chen, S., Wang, X.: Effects of sirt1
  silencing on viability, invasion and metastasis of human glioma cell lines.
  Oncology letters  \textbf{17}(4),  3701--3708 (2019)

\bibitem{lindstrom2019expanding}
Lindstr{\"o}m, M.S.: Expanding the scope of candidate prognostic marker igfbp2
  in glioblastoma. Bioscience reports  \textbf{39}(7),  BSR20190770 (2019)

\bibitem{rubanova2019latent}
Rubanova, Y., Chen, R.T., Duvenaud, D.: Latent odes for irregularly-sampled
  time series. In: Proceedings of the 33rd International Conference on Neural
  Information Processing Systems. pp. 5320--5330 (2019)

\bibitem{salhia2008guanine}
Salhia, B., Tran, N.L., Chan, A., Wolf, A., Nakada, M., Rutka, F., Ennis, M.,
  McDonough, W.S., Berens, M.E., Symons, M., et~al.: The guanine nucleotide
  exchange factors trio, ect2, and vav3 mediate the invasive behavior of
  glioblastoma. The American journal of pathology  \textbf{173}(6),  1828--1838
  (2008)

\bibitem{sato2020five}
Sato, K., Tahata, K., Akimoto, K.: Five genes associated with survival in
  patients with lower-grade gliomas were identified by information-theoretical
  analysis. Anticancer research  \textbf{40}(5),  2777--2785 (2020)

\bibitem{steponaitis2016high}
Steponaitis, G., Skiriut{\.e}, D., Kazlauskas, A., Golubickait{\.e}, I.,
  Stakaitis, R., Tama{\v{s}}auskas, A., Vaitkien{\.e}, P.: High chi3l1
  expression is associated with glioma patient survival. Diagnostic pathology
  \textbf{11}(1), ~1--8 (2016)

\bibitem{tomczak2015cancer}
Tomczak, K., Czerwi{\'n}ska, P., Wiznerowicz, M.: The cancer genome atlas
  (tcga): an immeasurable source of knowledge. Contemporary oncology
  \textbf{19}(1A), ~A68 (2015)

\bibitem{villani2009wasserstein}
Villani, C.: The wasserstein distances. In: Optimal Transport, pp. 93--111.
  Springer (2009)

\bibitem{yue2018high}
Yue, H., Xu, Q., Xie, S.: High emp3 expression might independently predict poor
  overall survival in glioblastoma and its expression is related to dna
  methylation. Medicine  \textbf{97}(1) (2018)

\bibitem{yildiz_ode2vae_2019}
Çağatay Yıldız, Heinonen, M., Lähdesmäki, H.: Ode$^2$vae: Deep generative
  second order odes with bayesian neural networks (2019)

\bibitem{zhang2019international}
Zhang, J., Bajari, R., Andric, D., Gerthoffert, F., Lepsa, A., Nahal-Bose, H.,
  Stein, L.D., Ferretti, V.: The international cancer genome consortium data
  portal. Nature biotechnology  \textbf{37}(4),  367--369 (2019)

\end{thebibliography}

\end{document}